\documentclass[twocolumn,times]{aastex63}

\received{XXXX XX, 2020}

\shorttitle{The Relation between the Optical \feii Emission and \mdot for AGN}
\shortauthors{Liu \& Bian}

\usepackage{graphicx}
\usepackage{multirow}
\usepackage{color}
\usepackage[figuresright]{rotating}
\usepackage{longtable}
\usepackage{mathrsfs,amsmath}
\usepackage{epstopdf}
\newsavebox{\tablebox}
\usepackage{hyperref}
\usepackage{times}
\usepackage{longtable}
\usepackage{array}
\usepackage{booktabs}

\usepackage{lineno}

\newcommand{\lv}{\ifmmode L_{5100} \else $L_{5100}$\ \fi}
\newcommand{\kms}{\ifmmode {\rm km\ s}^{-1} \else km s$^{-1}$\ \fi}
\newcommand{\ergs}{\ifmmode {\rm erg\ s}^{-1} \else erg s$^{-1}$\ \fi}
\newcommand{\lb}{\ifmmode L_{\rm Bol} \else $L_{\rm Bol}$\ \fi}
\newcommand{\ledd}{\ifmmode L_{\rm Edd} \else $L_{\rm Edd}$\ \fi}
\newcommand{\hb}{\ifmmode H\beta \else H$\beta$\ \fi}
\newcommand{\ha}{\ifmmode H\alpha \else H$\alpha$\ \fi}
\newcommand{\mgii}{Mg {\sc ii}}
\newcommand{\civ}{C {\sc iv}}
\newcommand{\oiii}{[O {\sc iii}]}

\newcommand{\feii}{Fe {\sc ii}\ }
\newcommand{\mbh}{\ifmmode M_{\rm BH}  \else $M_{\rm BH}$\ \fi}
\newcommand{\msun}{M_{\odot}}
\newcommand{\rfe}{\ifmmode R_{\rm Fe} \else $R_{\rm Fe}$\ \fi}
\newcommand{\sst}{\ifmmode \sigma_{\rm \ast}\else $\sigma_{\rm \ast}$\ \fi}
\newcommand{\dhb}{\ifmmode D_{\rm H\beta} \else $D_{\rm H\beta}$\ \fi}
\newcommand{\leddR}{\ifmmode L_{\rm Bol}/L_{\rm Edd} \else $L_{\rm Bol}/L_{\rm Edd}$\ \fi}
\newcommand{\mdot}{\ifmmode \dot{\mathscr{M}}  \else $\dot{\mathscr{M}}$\ \fi}
\newcommand{\rhb}{\ifmmode R_{\rm BLR}({\rm H\beta})  \else $R_{\rm BLR}({\rm H\beta})$ \ \fi}
\newcommand{\shb}{\ifmmode \sigma_{\rm H\beta} \else $\sigma_{\rm \hb}$\ \fi}
\newcommand{\RL}{\ifmmode R_{\rm BLR}({\rm H\beta}) - L_{\rm 5100} \else $R_{\rm BLR}({\rm H\beta}) - L_{\rm 5100}$ \ \fi}
\newcommand{\ms}{\ifmmode M_{\rm BH}-\sigma_{\ast} \else $M_{\rm BH}-\sigma_{\ast}$\ \fi}
\newcommand{\sm}{\ifmmode \sigma_{\rm H\beta,mean} \else $\sigma_{\rm H\beta,mean}$\ \fi}
\newcommand{\sr}{\ifmmode \sigma_{\rm H\beta,rms} \else $\sigma_{\rm H\beta,rms}$\ \fi}
\newcommand{\fwm}{\ifmmode \rm FWHM_{\rm mean} \else $\rm FWHM_{\rm mean}$\ \fi}
\newcommand{\fwr}{\ifmmode \rm FWHM_{\rm rms} \else $\rm FWHM_{\rm rms}$\ \fi}
\newcommand{\vpfm}{\ifmmode \rm VP_{\rm F, mean} \else $\rm VP_{\rm F, mean}$\ \fi}
\newcommand{\vpsm}{\ifmmode \rm VP_{\rm \sigma, mean} \else $\rm VP_{\rm \sigma, mean}$\ \fi}
\newcommand{\vpfr}{\ifmmode \rm VP_{\rm F, rms} \else $\rm VP_{\rm F, rms}$\ \fi}
\newcommand{\vpsr}{\ifmmode \rm VP_{\rm \sigma, rms} \else $\rm VP_{\rm \sigma, rms}$\ \fi}
\newcommand{\ffm}{\ifmmode f_{\rm F, mean} \else $f_{\rm F, mean}$\ \fi}
\newcommand{\fsm}{\ifmmode f_{\rm \sigma, mean} \else $f_{\rm \sigma, mean}$\ \fi}
\newcommand{\ffr}{\ifmmode f_{\rm F, rms} \else $f_{\rm F, rms}$\ \fi}
\newcommand{\fsr}{\ifmmode f_{\rm \sigma, rms} \else $f_{\rm \sigma, rms}$\ \fi}
\newcommand{\fc}{\ifmmode f_{\rm c} \else $f_{\rm c}$\ \fi}
\newcommand{\dhbm}{\ifmmode D_{\rm H\beta,mean} \else $D_{\rm H\beta,mean}$\ \fi}
\newcommand{\dhbr}{\ifmmode D_{\rm H\beta,rms} \else $D_{\rm H\beta,rms}$\ \fi}
\newcommand{\vfe}{\ifmmode v_{\rm Fe} \else $v_{\rm Fe}$\ \fi}

\newcommand{\lxm}{\ifmmode L_{\rm 14-195keV} \else  $L_{\rm 14-195keV}$ \fi}
\newcommand{\fx}{\ifmmode F_{\rm X} \else  $F_{\rm X}$ \fi}
\newcommand{\lx}{\ifmmode L_{\rm X} \else  $L_{\rm X}$ \fi}

\begin{document}

\title{The Relation between the Optical \feii Emission and the Dimensionless Accretion Rate for Active Galactic Nuclei}

\correspondingauthor{W. -H. Bian}
\email{whbian@njnu.edu.cn}

\author{Yan-Sheng Liu}
\affiliation{School of Physics and Technology, Nanjing
Normal University, Nanjing 210023, China}

\author{Wei-Hao Bian}
\affiliation{School of Physics and Technology, Nanjing
Normal University, Nanjing 210023, China}


\begin{abstract}
It was suggested that the prominent feature of the optical \feii emission  has a connection with the accretion process in active galactic nuclei (AGN). For a large sample of 4037 quasars ($z < 0.8$) with measured \hb line dispersion (\shb) selected from the Sloan Digital Sky Survey (SDSS) and 120 compiled reverberation-mapped (RM) AGN, we use \shb  and the extended \RL relation to calculate supermassive black holes masses (\mbh) from the single-epoch spectra for the SDSS subsample, and \shb from the mean spectra for the RM subsample.  
We find a strong correlation between the relative optical \feii strength \rfe and \mdot for the SDSS subsample with the Spearman correlation coefficient $r_s$ of $0.727$, which is consistent with that derived from the mean spectra for the RM subsample.  The magnitude of velocity shift of the optical \feii emission has a strong anticorrelation with \mdot, whenever there is inflow or outflow.
These strong correlations show that the optical \feii emission has an intimate connection with the accretion process.  Assuming that the difference of \mbh is due to the variable virial factor $f$ for adopting $ \rm FWHM_{\hb}$ as the velocity tracer,  we find that there is a relation between $f$ and $\rm FWHM_{\hb}$, $\log f=-(0.41\pm 0.002)  \rm \log FWHM_{\hb}+(1.719\pm 0.009)$ for the single-epoch spectrum.
The relation between $\log f$ and  \shb is not too strong, suggesting that \shb does not seem to depend much on the broad-line region inclination and a constant $\sigma$-based $f$ is suitable for \shb as the velocity tracer.

\end{abstract}



\section{Introduction} \label{sec:intro}
\feii emissions shown in the  optical and ultraviolet (UV) spectra are  prominent features  in most active galactic nuclei \citep[AGN; e.g.,][]{BG92, Hu2008, Shen2014, Zhao2020}. With the principal component analysis of a low-$z$ sample of 87 Palomar–Green quasars, \cite{BG92} found that principal component 1 (PC1) is related to the relative strength of optical \feii to the broad \hb (\rfe, the ratio between the strength of \feii emission within $4434-4684$ \AA~ and the broad \hb), the FWHM of the broad \hb ($\rm FWHM_{\hb}$), \oiii 5007 strength, and principal component 2 (PC2) links optical luminosity and $\alpha_{\rm ox}$ (optical-X-ray spectral index). 
PC1 and PC2 have relations with the accretion process around the supermassive black holes (SMBHs). With $\rm FWHM_{\hb}$-based SMBH masses (\mbh), \cite{Boroson2002} suggested that PC1 is mainly correlated with the Eddington ratio ($\leddR$, \lb is the bolometric luminosity and \ledd is the Eddington luminosity) and PC2 has a strong connection with \mbh and \leddR. 
Using a large sample of  the Sloan Digital Sky Survey (SDSS) quasars, \cite{Hu2008} did  a detailed line-fitting technique and found the connections between the optical \feii strength, velocity shift, and the Eddington ratio. 
Using a  two-dimensional plane of $\rm FWHM_{\hb}$ and \rfe for the SDSS quasars,  \cite{Shen2014} suggested that the Eddington ratio is also a key factor to be considered in the orientation-based unification of quasar phenomenology \citep{Urry1995}. 

In order to investigate the relation between \feii features and the accretion process, \mbh is a key parameter needing to be determined. With effort over nearly two decades, the reverberation-mapping (RM) method with the \hb spectral monitoring has  been successfully applied for about 120 AGN \citep[e.g.,][]{BM82, Pe04, Du2016a, Grier2017, Yu2020a, Yu2020b, Hu2021}. The RM method can measure the time lag between the  line variation ( e.g., \hb, \mgii, \civ)  and the corresponding continuum variation, and give the broad-line region (BLR) distance from the central SMBH ($R_{\rm BLR}$).  For type 1 AGN, the BLR clouds emitting the broad \hb line can be used as a probe to calculate the virial mass  \citep[e.g.,][]{Pe04, N2013}:
 \begin{equation}
 \label{eq1}
\mbh=f\times \frac{R_{\rm BLR}~(\Delta V)^2}{G}  .
\end{equation}
where  $\Delta V$ is the velocity of the BLR clouds,  $f$ is a corresponding virial factor, and $G$ is the gravitational constant.  Through the RM AGN, an empirical $R_{\rm BLR}-\lv$ relation ($L_{\rm 5100}$ is the 5100 \AA\ monochromatic luminosity, $R_{\rm BLR} \propto \lv^{0.5}$) was given and used to estimate $R_{\rm BLR}$ from a single-epoch spectrum  \cite[e.g.,][]{Ka00, Be13, Du2016b, Dalla2020,Khadka2022, Maithil2022}.  
The changes of the spectral energy distribution, connected to changes in the accretion rate, would likely lead to the breadth (an increment in the scatter) in the empirical $R_{\rm BLR}-\lv$ relation in the first-order consideration \citep{Wang2014, Ma2019, Du2019, Yu2020a, Maithil2022}. 
AGN with strong \feii emission show  a smaller  $R_{\rm BLR}$ than expected \citep[e.g.,][]{Du2019}. Since  \rfe has a relation with the accretion rate \citep[e.g.,][]{N2007,Du2016a, Yu2020a, Maithil2022},  an extended \RL relation ($R_{\rm BLR} \propto \lv^{0.5} + \rfe$) was recently suggested
\citep{Du2019, Yu2020a, Khadka2022}.   
The reduction in the scatter in the extended \RL relation \citep{Yu2020a} is in agreement with the one ($\sim 0.13$ dex) previously reported by \cite{KE15}, who used the UV luminosity instead of \lv as a substitute for the ionizing luminosity in the empirical \RL relation. The extended \RL relation would result in a better measurement of \mbh.


For the RM AGN, $\Delta V$ is  usually traced by four kinds of velocities, i.e.,  the broad \hb FWHM or the line dispersion ($\shb$) measured from the mean/rms spectrum \citep[e.g.,][]{Pe04, HK14, Yu2020b}. From Equation \ref{eq1}, $f=\mbh/(R_{\rm BLR} (\Delta V)^2/G)$.  For four kinds of velocities $\Delta V$, the corresponding four kinds of virial factors $f$ are defined, which are usually calibrated through the \ms relation ($\sigma_*$ is the bulge stellar velocity dispersion) or other independent methods to derive the SMBH masses \citep[e.g.,][]{On04, HK14, Yu2020b}. For several multiple RM AGN, it was found \shb is better than $\rm FWHM_{\hb}$ used to calculate \mbh with a constant factor $f$ \citep{Pe04}. For the 120 RM AGN sample \citep{Yu2020b}, with respect to \mbh from $\sigma_*$ or $\sigma_{\rm \hb,rms}$, it was found that we can obtain \mbh from $\sigma_{\rm \hb,mean}$ with the smallest scatter than from $\rm FWHM_{\hb}$. 
\cite{Dalla2020} also suggested that the use of \shb gives better results, while the use of $\rm FWHM_{\hb}$ introduces a bias, stretching the mass scale such that high masses are overestimated and low masses are underestimated, although both velocity tracers are usable.
A variable $\rm FWHM$-based $f$ was suggested \citep{Collin2006, Mejia2018, Yu2019, Yu2020b}. 
The cumulative fraction of $f$ was consistent with the  a simple model of thick-diskBLRs, which implied that, as the tracer of BLRs velocity, $\rm FWHM_{\hb}$ has some dependence on the BLRs inclination, while \shb is insensitive to the inclination \citep{Collin2006, Yu2019}.
For AGN with the single-epoch spectrum,  $\rm FWHM_{\hb}$ or \shb is used to trace $\Delta V$, which was used in large surveys, such as SDSS \citep{Sh11}.
With respect to the \hb FWHM, \shb was preferred to calculate \mbh from the single-epoch spectrum  \citep{Yu2020b}.  With Equation \ref{eq1}, $\mbh=f_{\sigma}\times \shb^2 \rhb/G=f_{\rm FWHM}\times {\rm FWHM}_{\hb}^2 \rhb/G$. 
Considering  a variable $f_{\rm FWHM}$ in $\rm FWHM_{\hb}$-based \mbh,  it means
 \begin{equation}
 \label{eq2}
f_{\rm FWHM}=f_{\sigma} (\shb / \rm FWHM_{\hb})^2.
\end{equation}

About the accretion strength, there are mainly two parameters, the Eddington ratio \lb/\ledd and the dimensionless accretion rate \mdot, where $\mdot\equiv \dot{M}/\dot{M}_{\rm Edd}$, $\dot{M}_{\rm Edd}=L_{\rm Edd}/c^2$.  The accretion rate $\dot{M}$ can be derived  from the disk model of \cite{SS73}, which has been extensively applied to fit the spectra of AGN \citep[e.g.,][]{Collin2002,Bian2003,  DL2011, Mejia2018}. Considering that the radius $R$ distribution of the effective disk temperature is given by $T_{\rm eff} \propto R^{-3/4}$,  \mdot is \citep[e.g.,][]{Du2016a}
 \begin{equation}
 \label{eq3}
\mdot \equiv \dot{M}/\dot{M}_{\rm Edd} = 20.1\left(\frac{l_{44}}{\cos\,\it{i}}\right)^{3/2} m_{7}^{-2}.
\end{equation}
where $l_{44} = \lv/10^{44}\ergs$ , $m_{7}=M_{\rm BH}/10^7 \msun$.  An average value of ${\rm cos}~ i = 0.75$ is adopted, which corresponds to the opening angle of the dusty torus \citep[e.g.,][]{DL2011, Du2016a}. \mdot is inversely proportional to \mbh to the second power.  This Equation applies to AGN that have $\log (\mdot/\msun)$ between -2  and 3.5, namely excluding the regimes of advection-dominated accretion flows ($\log (\mdot/\msun) <-2$) and of flows with hyperaccretion rates ($\log (\mdot/\msun) > 3.5$). 
Using FWHM-based \mbh and \lb/\ledd for 63 RM AGN, \cite{Du2016a} found \rfe correlated with \lb/\ledd and \mdot \citep[also see][]{N2007, Hu2008, Maithil2022}. They also suggested the BLRs "fundamental plane", i.e., a strong bivariate correlation of \mdot with \rfe and the \hb shape $\rm FWHM_{\hb}/\shb$.

In this paper, we use a large sample of 4037 SDSS quasars \citep[$z<0.8$;][]{Hu2008} with measured broad \hb line dispersion \shb and \rfe to investigate again the relation between \feii emission and the accretion process, as well as a subsample of 120 RM AGN. About the accretion strength, we use \mdot instead of \leddR. We also derive the relation between $f_{\rm FWHM}$ and $\rm FWHM_{\hb}$ for the single-epoch spectrum. This paper is organized as follows. Section 2 presents the adopted samples and data analysis. Section 3 is our results and discussions. Section 4 is our conclusions. All of the cosmological calculations in this paper assume $\Omega_{\Lambda}=0.7$, $\Omega_{\rm M}=0.3$, and $H_{0}=70~ \kms {\rm Mpc}^{-1}$.


\section{Samples and data Analysis}
\begin{figure}
\includegraphics[angle=0,width=3.4in]{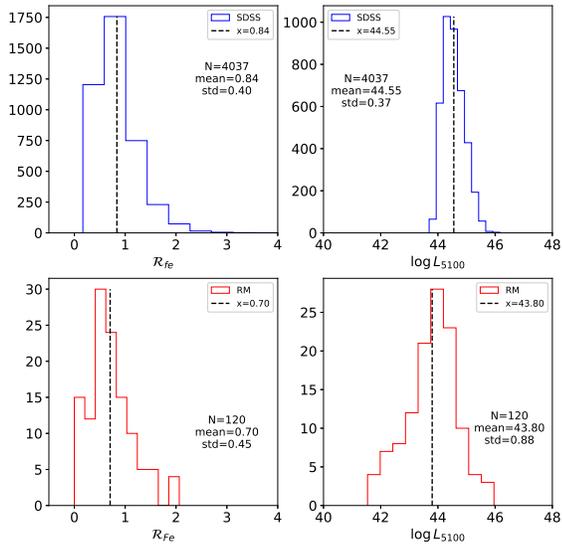}
\caption{Distributions of \rfe  and \lv for the subsample of  4037 SDSS quasars (top panels, blue) and the subsample of 120 RM AGN (bottom panels, red). The dashed lines are their mean values. The number, the mean value, and the standard deviation are shown in panels. }
\label{fig1}
\end{figure}

\begin{figure*}
\center
\includegraphics[angle=0,width=6.6in]{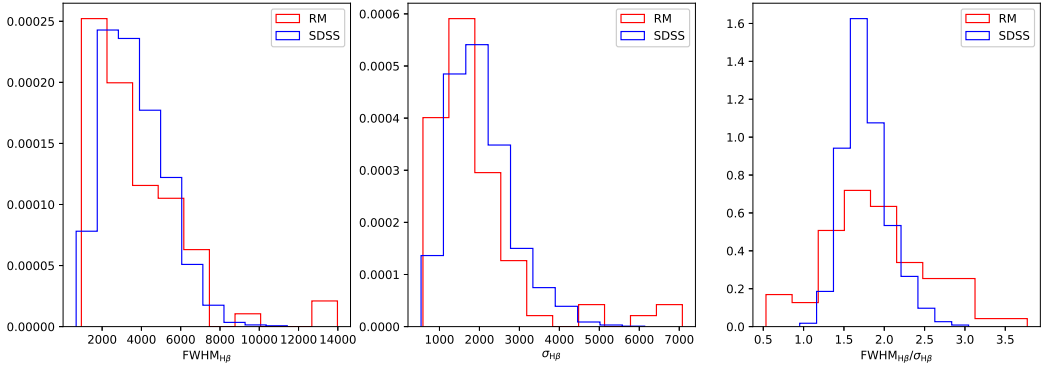}
\caption{Distributions of $\rm FWHM_{\hb}$, \shb, and their ratio $\rm FWHM_{\hb}/\shb$ for the broad \hb.  The blue lines are for the subsample of 4037 SDSS quasars. The red lines are for the subsample of 120 RM AGN. The fractional numbers are used to compare the distributions for two subsamples.}
\label{fig2}
\end{figure*}

Because we use \shb instead of $\rm FWHM_{\hb}$ to calculate \mbh, the spectral decomposition is crucial. We use a subsample of 4037 quasars ($z<0.8$)  from SDSS presented by \cite{Hu2008} with  criteria of signal-to-noise ratio $(\rm S/N)>10$ and $\rm EW (Fe~II) > 25~ \AA$.  We briefly introduce their spectral fitting \citep{Hu2008}. The continuum is decomposed into three components: a single power law, Balmer continuum emission, and a pseudocontinuum due to blended \feii emission. For the pure emission-line spectrum subtracting the continuum, the narrow \hb component, \oiii 4959, 5007 are modeled using Gaussians, and the broad \hb component is modeled using a Gauss-Hermite function, whose best fit yields the $\rm FWHM_{\hb}$, \shb. At the same time, we also use a compiled subsample of 120 RM AGN with measured four kinds of velocity tracers from mean and rms spectra \citep[][and references therein]{Yu2020b}.  About the 120 RM subsample \citep{Yu2020a}, we mainly adopted the data from \cite{Du2016a} and \cite{Shen2019}.  For 63 AGN in this RM subsample, \cite{Du2016a} used the same fitting scheme as \cite{Hu2008}. They included the host correction in the fitting described in \cite{Hu2015}. For other 44 SDSS AGN in the RM subsample, \cite{Shen2019} used almost the same fitting scheme as \cite{Hu2015}.  For this subsample of 120 RM AGN, the flux of broad optical \feii is all measured by integration from 4434 to 4684 \AA, $z$ is less than 1.026, host-corrected \lv is between $10^{41.54}$ and $10^{45.95}$ \ergs. In order to use all of available RM AGN, no criterion on $\rm EW(Fe~II)$ or  $\rm S/N$ is used. 

For the SDSS subsample, \lv,  the \feii luminosity,  the broad \hb luminosity, $\rm FWHM_{\hb}$ and $\rm \sigma_{\hb}$  are adopted, respectively, from columns (15), (3), (9), (11), (41)  in Table 2 in \cite{Hu2008}.  The ratio of the luminosity of \feii to the broad \hb is calculated as \rfe with its error calculated from  the error transfer formula. With respect to \oiii ~$\lambda$5007, the \feii velocity shift and its error are adopted from columns (7)-(8) in Table 2 in \cite{Hu2008}. For the RM subsample, \lv and \rfe are adopted frm columns (3) and (7)  in Table 1 in \cite{Yu2020a}, as well as their errors. 
Using the relative variance fraction $F_{\rm var}$, \cite{Hu2015} gave \feii RM for 10 narrow-line Seyfert 1 galaxies (NLS1s). They found that all 10 objects show \feii variations with an amplitude of a few to 10\%. On average, the variability of \feii is about 10\% smaller than the variability of the \hb line.  As the variability of \hb is unusually much larger than that of \feii, the uncertainties of \rfe are mainly governed by \hb variability, which on average is $\sim 20\%$ \citep{Hu2015, Du2016a}. \fwm, \sm, \fwr, \sr and their errors are adopted from columns (4)-(7) in Table 1 in \cite{Yu2020b}.  

In Figure \ref{fig1}, we show the distributions of \lv and \rfe for the SDSS subsample and the RM subsample.  It is clear that the RM subsample has averagely smaller \lv than the  SDSS subsample. 
We perform the Kolmogorov– Smirnov (K–S) test \footnote{We used python (scipy.stats.ks\_2samp, scipy.stats.spearmanr) to do our following analysis for the  K-S test and the Spearman correlation test. } on distributions in Figure \ref{fig1}. The statistic d and the significant level probability for the null hypothesis 
($p$-value) are -0.18 and 0.0001 for \rfe,  0.52 and $2.22\times10^{-16}$ for \lv, respectively.  The distribution of \rfe has a difference between the two subsamples and the difference of the \lv distribution is more significant. 
In Figure \ref{fig2}, we show the distributions of $\rm FWHM_{\hb}$, \shb, and the broad \hb shape $\dhbm \equiv \rm FWHM_{\hb}/\shb$ from the mean spectrum for the RM subsample (red lines).  It was found that $\dhbm$ has a relation with \leddR or \mdot \citep[e.g.,][]{Collin2006, Du2016a, Yu2020a, Panda2022}. 
We also show the distribution of $\rm FWHM_{\hb}$ and \shb from the single-epoch spectrum for the SDSS subsample (blue lines).
We perform the K–S test on distributions in Figure \ref{fig2} and find no significant differences between the two samples. The statistic d and p-value for the null hypothesis are 0.17 and 0.03 for $\rm FWHM_{\hb}$, 0.20 and 0.005 for \shb, 0.24 and 0.0004 for $\rm FWHM_{\hb}/\shb$, respectively.  The value of $\rm FWHM_{\hb}/\shb$ is 2.35, 3.46, 2.45, 2.83, and 0 for a Gaussian, a rectangular, a triangular, an edge-on rotating ring, and a Lorentzian profile, respectively \citep{Collin2006, Du2016a}. 
For the SDSS subsample,  the mean value and its standard deviation for $\rm FWHM_{\hb}/\shb$  are $1.78$ and $0.28$ (see right panel in Figure \ref{fig2}).  For the RM subsamples, the mean value and its standard deviation are $1.94$ and $0.61$. They show the deviation from the value of $2.35$ for the single Gaussian profile. 
There are 37 NLS1s with $\rm FWHM_{\hb} < 2000~ \kms$ from the mean spectrum for the RM subsample and 564 NLS1s  from the single-epoch spectrum for the SDSS subsample \citep[e.g.,][]{Bian2004}.  With respect to the proportion of the number of NLS1s in the SDSS subsample ($\sim 14\%$), the higher proportion in the RM subsample ($\sim 31\%$) is due to including an RM project searching for super-Eddington accreting massive black holes \citep[SEAMBHs;][]{Wang2013, Du2016a}.  The mean and its standard deviation of $\dhbm$ for NLS1s are $1.53$ and  $0.27$ for the RM subsample, $1.50$ and $0.15$ for  the SDSS subsample, which are much deviated from the value of $2.35$ for the single Gaussian profile.

For the SDSS subsample, \lv is from $\sim 10^{44}$ to $\sim 10^{46}$ \ergs, the mean value and the standard deviation of $\log \lv$ are  44.55 and 0.37. We use the empirical formula to correct the host contribution \citep{Sh11, Ge2016},
 \begin{equation}
 \label{eq4}
\frac{L_{\rm 5100, host}}{L_{\rm 5100,QSO}}=0.8052-1.5502x+0.9121x^2-0.1577x^3.
\end{equation}
for $x + 44 \equiv \log(L_{\rm 5100,total} /\ergs ) < 45.053$. There are 3590/4037 AGN with luminosities below this value, and no correction is needed for luminosities above this value.  For host-corrected $\log (\lv/\ergs)$, the mean value is 44.40, which is smaller than uncorrected mean values of 44.55 by 0.15 dex.

\section{Results and Discussions}

\subsection{\mbh, \mdot and \lb/\ledd}
\begin{figure}
\includegraphics[angle=0,width=3.4in]{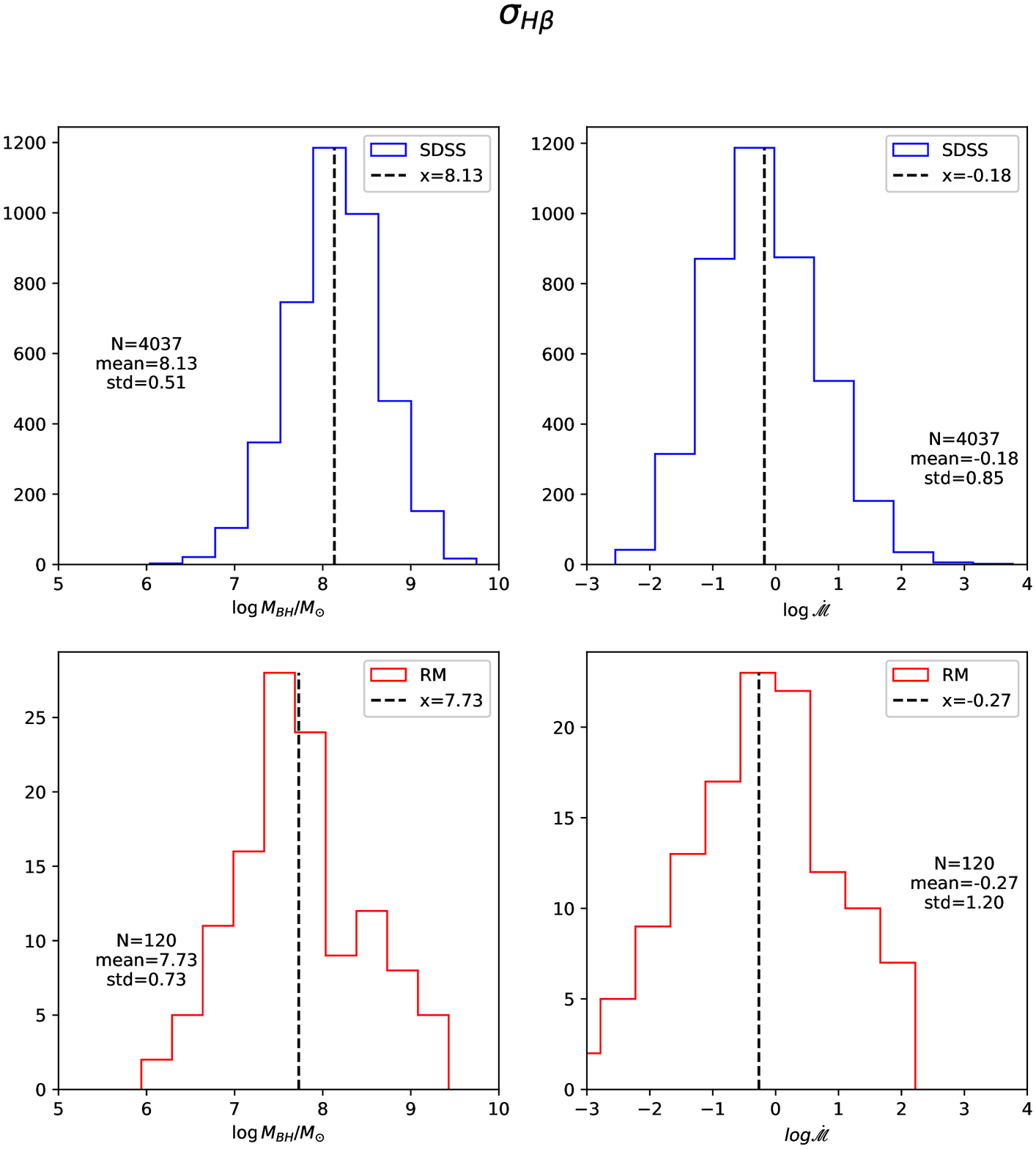}
\includegraphics[angle=0,width=3.4in]{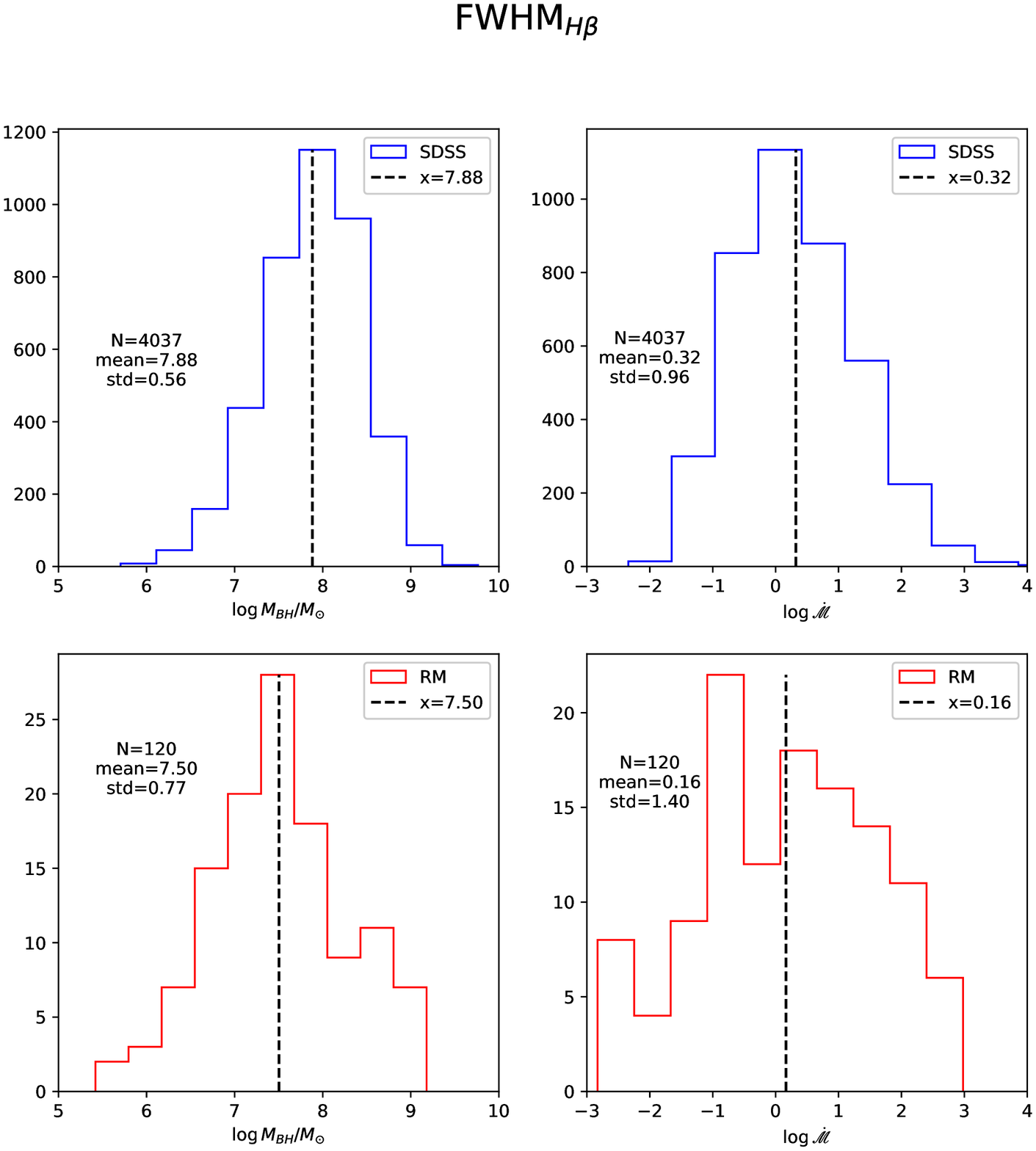}
\caption{Distributions for  \mbh and \mdot derived from \shb (top two panels) and $\rm FWHM_{\hb}$ (bottom two panels) for the subsample of  4037 SDSS quasars (blue) and the subsample of 120 RM AGN (red). The dashed lines are mean values. The number, the mean value, and the standard deviation are shown in panels.}
\label{fig3}
\end{figure}

\begin{figure}
\includegraphics[angle=0,width=3.4in]{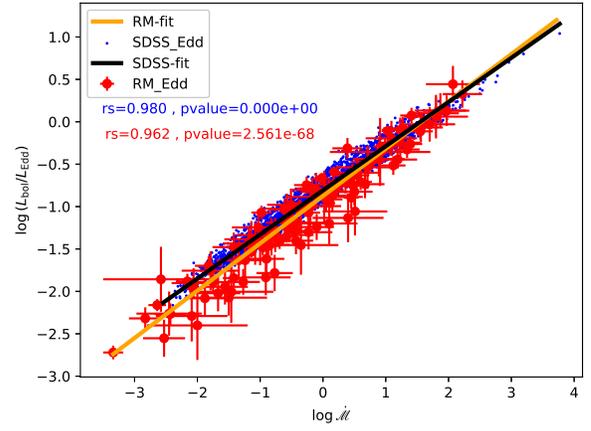}
\caption{\leddR versus \mdot   for the subsample of  4037 SDSS quasars (blue points) and the subsample of 120 RM AGN (red points). The black and orange solid lines are the best linear fittings for the SDSS subsample and the RM subsample. }
\label{fig4}
\end{figure}

An extended empirical \RL relation including \rfe was found for the RM AGN subsample \citep{Du2019, Yu2020a}:
\begin{equation}
\label{eq5}
\begin{split}
\log \frac{\rhb}{ltd}=(0.48\pm0.03)\log l_{44}
-(0.38\pm0.04)\rfe\\+(1.67\pm0.09)
\end{split}
\end{equation}
For $\log \rhb$, it has an intrinsic scatter of 0.17 dex \citep{Yu2020a}. We adopt a systematic error of 0.2 dex in our following calculation. 
With the extended \RL relation, when \shb is used as the tracer of the virial velocity, and $f$ is adopted as 5.5 \citep[e.g.,][]{Yu2020a, Yu2020b},
the single-epoch virial mass formula can be expressed as
\begin{equation}
\label{eq6}
\begin{split}
\log \frac {M_{\rm BH,\shb}}{M_{\odot}} =7.7+2\log \frac{\rm \sigma_{H\rm \beta}}{\rm 1000~ \kms} & \\
+(0.48\pm 0.03)\log l_{44} -(0.38\pm 0.04)\rfe
\end{split}
\end{equation}
When we use $\rm FWHM_{\hb}$ as the tracer of the virial velocity, and adopt $f$ as 1 \citep[e.g.,][]{Du2019},
the single-epoch virial mass formula can be expressed as
\begin{equation}
\label{eq7}
\begin{split}
\log \frac {M_{\rm BH,FWHM_{\hb}}}{M_{\odot}} =6.96+2\log \frac{\rm FWHM_{\hb}}{\rm 1000~ \kms} & \\ 
+(0.48\pm 0.03)\log l_{44} -(0.38\pm 0.04)\rfe
\end{split}
\end{equation}
Substitute Equation \ref{eq6} for Equation \ref{eq3}, the \mdot formula is:
\begin{equation}
\label{eq8}
\begin{split}
\log (\mdot_{\shb}) = (0.54\pm 0.06) \log l_{44}-4\log \frac{\shb}{\rm 1000~ \kms} & \\
+(0.76\pm 0.08)\rfe+0.09 \\
\end{split}
\end{equation}
When using $\rm FWHM_{\hb}$, a similar \mdot formula is:
\begin{equation}
\label{eq9}
\begin{split}
\log (\mdot_{\rm FWHM_{\hb}}) = (0.54\pm 0.06) \log l_{44}-4\log \frac{\rm FWHM_{H\rm \beta}}{\rm 1000~ \kms} & \\
+(0.76\pm 0.08)\rfe+1.57 \\
\end{split}
\end{equation}
We use a correction factor to calculate \lb from \lv, $\lb= BC_{\rm 5100} \lv$, where  $BC_{\rm 5100}=53-\log (\lv/\ergs)$ \citep{Marconi2004, N2013}.  In the following sections,  if not stated, $\rm FWHM_{\hb}$ and \shb are from the mean spectrum for the RM subsample, and we will adopt the $\shb$-based \mbh and \mdot from the single-epoch spectrum for the SDSS subsample and  from the mean spectrum for the RM subsample. 

For the SDSS subsample of 4037 AGN, using \shb or $\rm FWHM_{\hb}$ from the single-epoch spectrum and the extended \RL relation, from Equations \ref{eq6} or \ref{eq7}, we calculate \mbh. Based on the error transfer formula , the error of \mbh is calculated from the errors of \shb or $\rm FWHM_{\hb}$, \lv, \rfe, including a systematic error of 0.2 dex for \rhb.  For the RM subsample of 120 RM AGN, using \shb or $\rm FWHM_{\hb}$ from the mean spectrum and \rhb measured through RM technique, from Equation \ref{eq1},  \mbh was presented in \cite{Yu2020b}, as well as its error.  We use Equation \ref{eq3} to calculate \mdot, as well as its error.
In Figure \ref{fig3}, we show the distributions of \mbh (left panels) and \mdot (right panels) for the SDSS subsample and the RM subsample. Left two top panels show the mean value  and  the standard deviation of $\log \mbh$ from \shb are 8.13 and  0.51 for the SDSS subsample; 7.73 and 0.73 for the  RM subsample. The former is averagely larger than the latter by 0.4 dex. It is mainly due to a large \lv for the SDSS subsample. The left two bottom panels show that for \mbh calculated from $\rm FWHM_{\hb}$. For the SDSS subsample , the mean values of $\shb$-based \mbh  and $\rm FWHM_{\hb}$-based \mbh are 8.13 and 7.88.  For the RM subsample , the mean values of $\shb$-based \mbh  and $\rm FWHM_{\hb}$-based \mbh are 7.73 and 7.5. The $\shb$-based \mbh is averagely larger than $\rm FWHM_{\hb}$-based \mbh for the SDSS subsample (by 0.25 dex) or the RM subsample (by 0.23 dex).  It is smaller than $0.5$ dex found in \cite{Bian2008}.  The right two top panels show \mdot calculated from \shb. Right two bottom panels shows that for \mdot calculate from $\rm FWHM_{\hb}$.  The $\shb$-based \mdot is averagely smaller than $\rm FWHM_{\hb}$-based \mdot for the SDSS subsample (by 0.5 dex) or the RM subsample (by 0.43 dex).  
We perform the K-S tests on the distributions in Figure  \ref{fig3} between the SDSS subsample and the RM subsample.
For $\shb$-based \mbh, $\rm d=0.344$, p-value=$8.60\times10^{-13}$; for $\shb$-based \mdot, $\rm d=0.123$, p-value=$0.053$; 
for $\rm FWHM_{\hb}$-based \mbh, $\rm d=0.320$, p-value=$4.29 \times10^{-11}$; for $\rm FWHM_{\hb}$-based \mdot, $\rm d=0.165$, p-value=$0.0031$.
The difference of \mdot for these two subsamples is not significant while the \mbh difference for these two subsamples is significant.

A criterion of $\mdot \geq 3$ is adopted to select  super-Eddington accreting massive black holes  \citep[e.g.,][]{Wang2013, Du2016a}. Using $\shb$-based $\mdot$,  there are 912/4037 super-Eddington accreting AGN in the SDSS subsample, and 40/120 super-Eddington accreting AGN in the RM subsample. A relative large proportion of  super-Eddington accreting AGN  in the SDSS subsample is due to the criterion of $\rm EW(Fe II) > 25 \AA$.  For the RM subsample, there are 25 AGN presented by the SEAMBH collaboration \citep{Wang2013, Du2016a}.

In Figure \ref{fig4}, we show the \leddR and \mdot  for the SDSS subsample  (blue dots) or the RM subsample (red dots). For the SDSS subsample, considering the errors in both coordinates, the best linear fitting through \textsf{kmpfit} \footnote{https://www.astro.rug.nl/software/kapteyn/kmpfittutorial.html. \textsf{kmpfit} is the Kapteyn Package Python binding for a piece of software that provides a robust and relatively fast way to perform nonlinear least-squares curve and surface fitting.} is $\log \leddR=-(0.808\pm 0.001)+(0.521\pm 0.002)\log \mdot$. For the RM subsample, the best linear fitting is $\log \leddR=-(0.876\pm 0.013)+(0.558\pm 0.013)\log \mdot$. The latter relation is steeper than the former. The slope difference is due to the larger \lv for the SDSS subsample (see Figure \ref{fig1}). It is consistent with the result by \cite{Maithil2022} (see their Figure 6). These slopes are consistent with the slopes of $0.53$ found by \cite{DL2011} and $0.52$ by \cite{Huang2020}. 
Using Equations \ref{eq1} and \ref{eq3}, $\mdot \propto \lv^{1.5}\mbh^{-2}$, $\leddR \propto \lb \mbh^{-1} \propto \lv \mbh^{-1}$,  so $\leddR \propto \mdot^{0.5} \lv^{0.25}$. The dependence on \lv is smaller than on \mdot. The slope of $0.5$ is well consistent with our slope of $0.52$ or  $0.56$.  The nonlinear relation between \leddR and \mdot  (although it is linear when the values of the logarithm are considered) would have an impact on the relations using \mdot instead of \leddR. Considering $\lb=\eta \dot{M}c^2$, where $\eta$ is the accretion efficiency, $\leddR=\frac{\eta \dot{M}c^2}{\dot{M}_{\rm Edd}c^2}=\eta \mdot$.  Adopting $\leddR \propto \mdot ^{0.5}  \lv^{0.25}$, it means that $\eta \propto \mdot^{-0.5} \lv^{0.25} \propto (\leddR)^{-1} \lv^{0.5}$. The nonlinear relation implies that the converting efficiency $\eta$ is anticorrelated with \mdot or \leddR ( including a relatively weaker dependence on \lv), i.e., smaller efficiency with higher dimensionless accretion rate. 

\subsection{\rfe versus \mdot}
\begin{figure*}
\center
\includegraphics[angle=0,width=6.5in]{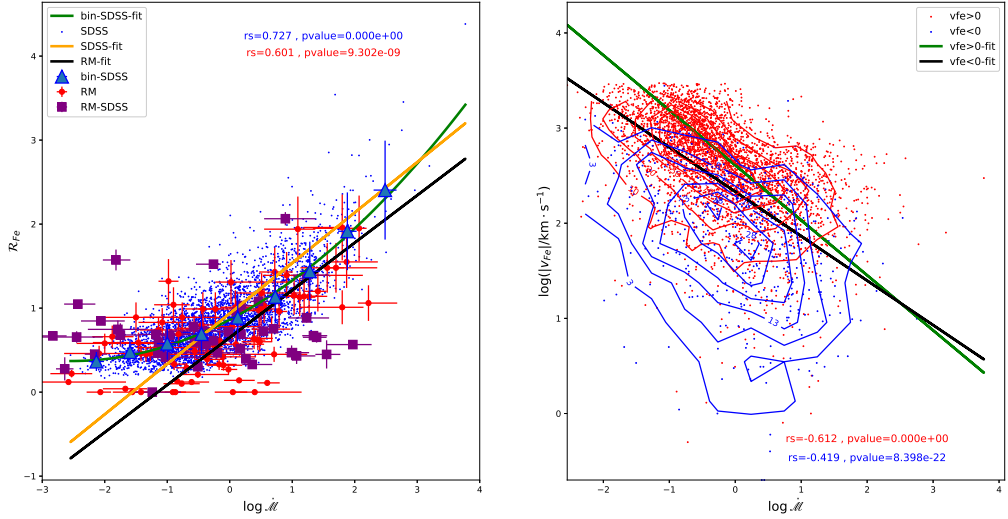}
\caption{Left: \rfe versus \mdot. The green line and black line are the best linear fittings for the SDSS subsample (blue points) and the RM subsample (red points). The blue triangles represent the mean \rfe in bins of $\Delta \log \mdot = 0.6 ~\rm dex$, and the error bars represent the standard deviations.  The green solid line shows the best fit to the binned data with a second-order polynomial function. The purple squares denote 44 RM SDSS AGN from the RM subsample. The $r_s$ and $p_{\rm null}$ for the RM subsample are the ones excluding the SDSS RM AGN. Right: \vfe versus \mdot for the SDSS subsample. The green line and black line are the best linear fittings for the inflow AGN ($\vfe >0$, red points and red contours) and the outflow AGN ($\vfe<0$, blue points and blue contours). }
\label{fig5}
\end{figure*}
In Figure \ref{fig5}, we show the relation between \rfe and \mdot. For the RM subsample, the Spearman correlation test gives a strong correlation with the coefficient $r_s=0.423$ and the probability of the null hypothesis $p_{\rm null}=1.48\times 10^{-6}$. Considering the errors in both coordinates, the kmpfit fitting gives $\rfe=(0.713\pm0.047)+(0.532\pm 0.066)\log \mdot$. Excluding 44 SDSS RM AGN from the subsample of 120 RM AGN, the correlation is stronger with  $r_s=0.60, p_{\rm null}=9.3\times 10^{-9}$. Considering the errors in both coordinates, the fitting gives $\rfe=(0.548\pm0.053)+(0.564\pm 0.074)\log \mdot$. Considering errors, these slopes are consistent. For the SDSS subsample, the correlation is strong with $r_s=0.727, p_{\rm null}=0$.  Considering the errors in both coordinates, the best linear fitting through kmpfit gives $\rfe=-(0.936\pm0.006)+(0.600\pm 0.008)\log \mdot$.  There exists some difference in intercepts for the SDSS subsample and the RM subsample. The slightly intercept difference implies that $\shb$-based \mbh for the SDSS subsample is slightly overestimated leading to an underestimated in \mdot.  
For 44 SDSS RM AGN from the RM subsample, the correlation between  \rfe and \mdot is very weak with $r_s=0.06, p_{\rm null}=0.686$. 
This weak correlation is possibly due to the large uncertainties in the RM measurement of $R_{\rm BLR}$ for SDSS-RM AGN.
Excluding these 44 SDSS RM AGN from the RM subsample, for the relation between \rfe and \mdot, $r_s$ increases from 0.423 to 0.60, and   $p_{\rm null}$ decreases from $1.48\times 10^{-6}$ to  $9.3\times 10^{-9}$. Considering the errors in both coordinates, for these 44 SDSS RM AGN from the RM subsample, the kmpfit fitting gives $\rfe=(0.661\pm0.072)+(0.466\pm 0.115)\log \mdot$.  Although with a large scatter, its slope of $0.466$ is flatter than for the fitting relation  for the SDSS subsample ($0.6$) and the total RM subsample ($0.532$).  The underestimation of \mdot is smaller than for the SDSS subsample. 
We recognize that the much larger correlation coefficient ($r_s=0.727$ for the SDSS subsample) is in part an effect of self-correlation induced by \rfe in the extended R-L relationship to estimate \mbh and \mdot. Including \rfe to calculate \mdot, the distribution of the SDSS subsample is well consistent with that for the RM subsample (see left panel in Figure \ref{fig5}).  

We investigate this correlation for super-Eddington and sub-Eddington AGN (see Table \ref{tab1}). For 912 super-Eddington AGN in the SDSS subsample \citep[$\mdot >3$; e.g.,][]{Du2016a}, $r_s=0.57$  and $p_{\rm null}= 2.35\times 10^{-79}$. For 3125 sub-Eddington AGN  in the SDSS subsample,  $r_s=0.56$ and $p_{\rm pull}=3.45\times 10^{-252}$. For super-Eddington accreting AGN, the slope is steeper than for sub-Eddington accretion AGN. For the RM subsample, for 40 super-Eddington RM AGN, $r_s=0.43$  and $p_{\rm null}= 0.0052$.  For 80 sub-Eddington RM AGN,  $r_s=0.1$  and $p_{\rm null}= 0.38$, which shows the correlation is not significant.

\begin{table*}
\caption{The Spearman correlation coefficient and the probability of the null hypothesis for relations.}
\centering
\label{tab1}
\begin{lrbox}{\tablebox}
\begin{tabular}{l |l |l | l|l|l|l}
\hline
            & \multicolumn{3}{c|}{SDSS Subsample}   & \multicolumn{3}{c}{RM Subsample} \\  \cline{2-7}                   
            & All & $\mdot <3$ & $\mdot \geq 3$ & All & $\mdot <3$ & $\mdot \geq 3$\\
\hline
Number         &  4037 & 3125 & 912 & 120 & 80 & 40 \\     
\hline
\rfe-\mdot     & (0.727, 0)       &  (0.56, $3.45\times 10^{-252}$)      &    (0.57, $2.35\times 10^{-79} $ )      &   (0.423, $1.48\times10^{-6}$)  &  (0.1, 0.38)     & (0.43, 0.0052)                \\                     
$\vfe-\mdot ^*$    & (-0.612,0)$^*$      &  (0.52, $2.37\times 10^{-190}$)$^*$      &    (-0.08, 0.02 )  & -- & -- &--    \\    
\leddR-\mdot & (0.98, 0)        &    --       &     --         &  (0.962, $2.56\times10^{-68}$)  & --         & --           \\             
\hline  
\end{tabular}
\end{lrbox}
\scalebox{0.85}{\usebox{\tablebox}}
\\Note: The first value in brackets is the Spearman correlation coefficient and the second value is the probability of the null hypothesis. \\
*: For the  3556 AGN with \feii inflow ($\vfe >0$) in the SDSS subsample with 2809 sub-Eddington ($\mdot <3$) and 747 super-Eddington ($\mdot \geq 3$).
\label{tab1}
\end{table*}

The systematic uncertainty is about $0.3 ~\rm dex$ in \mbh, $0.6~\rm dex$ in \mdot \cite[e.g., ][]{Yu2020b}. For the relation between \rfe and \mdot, using $\Delta \log \mdot=0.6 ~\rm dex$ as the bin width, we calculate the mean and the standard deviation as the error in each bin.  It shows a steeper slope with increasing \mdot. Using 2th-order polynomial function, we find the best fitting is $\rfe=(0.863\pm 0.024)+(0.389\pm 0.014)\log \mdot+(0.077\pm 0.009)(\log \mdot)^2$, which can be used to estimate  \mdot from \rfe \citep{Du2016a}.

For a slightly smaller RM sample of 63 AGN \citep{Du2016a},  using FWHM-based \mdot, it was found that $\rfe=(0.66\pm 0.04)+(0.30\pm 0.03) \log \mdot$ ($r_s=0.65$). 
For 9818 SDSS DR4 AGN, it was found that, using FWHM-based \mbh, there was a correlation between \rfe  and  \leddR shown in Figure 5 of \cite{N2007}, $\rfe \propto (\leddR)^{0.7}$. Considering the nonlinear relation between \leddR and \mdot, $\leddR \propto \mdot^ {0.52}$, it implied $\rfe \propto \mdot^ {0.36}$ for 9818 AGN by  \cite{N2007}, which is consistent with the result by \cite{Du2016a}.  These slopes are flatter than ours, which is due to larger \mbh and  smaller \mdot using \shb instead of \shb. 
The strong correlation between \rfe and \mdot or \leddR show that the relative strength of optical \feii is driven by the strength of the accretion process. We also can use \rfe to trace  \mdot or \leddR though with a large scatter.  
Including the \hb shape $\rm FWHM_{\hb}/\shb$, for the SDSS subsample, the $r_s$ increases from 0.727 to 0.75. For the RM subsample, the $r_s$ increases from 0.423 to 0.46; $p_{\rm null}$ decreases from $1.48\times 10^{-6}$ to $1.12\times 10^{-7}$. It is found that, including  $\rm FWHM_{\hb}/\shb$, we cannot improve significantly the relation between \mdot and \rfe \citep[see also][]{Du2016a}. 

\subsection{\vfe versus \mdot}
For the SDSS subsample of 4037 AGN, a systematic redshift of optical \feii relative to \oiii$\lambda 5007$  (\vfe) was found by \cite{Hu2008}. 
The errors of \vfe were calculated from the fitting of the continuum and the fitting of the \oiii\ line, which was consistent with that given by the simulations. The median of the \vfe distribution is 116 \kms \citep{Hu2008}.
\cite{Sulentic2012} called into question the measurement of \vfe in \cite{Hu2008} .  The contradictory results are possibly due to the different ways to construct their composite spectra.  \cite{Sulentic2012} constructed them from bins by parameters of their 4D  Eigenvector 1 (EV1) formalism, namely, \hb FWHM and \rfe, while \cite{Hu2008} constructed them from bins in \vfe.  \cite{Hu2012} confirmed that the redshift measurements of \feii are robust.

For the SDSS subsample of 4037 AGN, it is found there is a relation between \vfe and $\rm FWHM_{Fe II}$ with $r_s=0.59, p_{\rm null} =0$. For the relation between \vfe and $\rm FWHM_{\hb}$, $r_s=0.50, p_{\rm null} =9.51\times 10^{-229}$ \citep[also see Figure 9 in ][]{Hu2008}.  It was found that the Eddington ratio is the main physical driver for \vfe \citep{Hu2008}. In the right panel in Figure \ref{fig5}, for the SDSS subsample of 4037 AGN, we show the relation between \vfe and \mdot. The correlations are strong with $r_s=-0.612, p_{\rm null}=0$ for 3556 AGN with \feii inflow ($\vfe >0$),  and $r_s=-0.419, p_{\rm null}=9.40\times 10^{-22}$  for 480 AGN with \feii outflow ($\vfe <0$), respectively. Considering the errors in both coordinates, the kmpfit fitting shows that $\log \vfe=(2.610\pm 0.007)+(-0.578\pm 0.01)\log \mdot$ for $\vfe >0$ (green line in right panel in Figure \ref{fig5}). It is $\log |\vfe |=(2.331\pm 0.018)+(-0.467\pm 0.022)\log \mdot$ for  $\vfe <0$ (black line).  
For 3556 AGN with \feii inflow ($\vfe >0$)  in the SDSS subsample, there are 2809 sub-Eddington AGN ($\mdot < 3$), $r_s=-0.52$  and $p_{\rm null}= 2.37\times 10^{-190}$. For 747 super-Eddington AGN  with \feii inflow ($\vfe >0$), $r_s=-0.08$, and $p_{\rm null}=0.02$, which shows the correlation is not significant (see Table \ref{tab1}).

Using \shb and $\rhb\propto \lv^{0.69}$ \citep{McGill2008}, for \feii inflow ($\vfe >0$),  it was found that $\log \vfe=(1.00\pm 0.05)+(-1.83\pm 0.05)\log \leddR$, $r_s=-0.53, p_{\rm null} <1.0\times 10^{-5}$ \citep{Hu2008}. Considering the nonlinear relation between \leddR and \mdot, $\leddR \propto \mdot^ {0.52}$, it implied $\vfe \propto \mdot^ {-0.95}$. Here using \shb, the extended \RL relation including \rfe and the host-corrected \lv to calculate \mbh, and the standard accretion disk to calculate \mdot,  we find that the slope is $-0.578\pm 0.01$. The flatter relation is due to the smaller \rhb for larger \rfe, which leads to smaller \mbh and larger \mdot.  
For the relation between \vfe and \mdot shown in right panel in Fig. 5, excluding extreme accretors with log
$\mdot$ larger than 1.5, we found that the correlation does not improve. It becomes slightly weaker with $r_s$ changing from 0.612 to 0.602.
For  \feii outflow ($\vfe <0$) , we also find a flatter relation with a slope of $-0.467\pm 0.022$ (see the black line in the right panel in Figure \ref{fig5}).  It suggests that the inflow velocity is slightly larger than the outflow velocity for AGN with the same \mdot.
The velocity of \feii is  driven by gravity toward the center and decelerated by the radiation pressure, as well as its initial velocity.  The redward shift of \feii and the inverse correlation between \vfe and \mdot favor a scenario in which \feii emission emerges from an inflow \citep{Hu2008}. An increase in \mdot would enhance the radiation pressure and lead to a decrease of the inward velocity of the inflow. 

From the general form of the equation of motion for a cloud of mass $M_c$, the acceleration $a(r)=a_{\rm rad}(r)-g(r)-\frac{1}{\rho}\frac{dP}{dr}+f_{d}/M_{c}$ \citep{N2013}, where $g(r)$ is the gravitational acceleration, $a_{\rm rad}(r)$ is the acceleration due to radiation pressure force, and $f_d$ is the drag force; $P$ is the thermal pressure [$P=\rho v_s^2$, where $v_s=(2k_BT/\mu m_{\rm H})^{1/2}\sim 16.6(T/10^4 K)^{1/2} \kms$, $\mu m_{\rm H}$ is the average mass per particle, and $\mu= 0.6$]. Neglecting  the gas pressure gradient and the drag force for pure wind flows, $a(r)=a_{\rm rad}(r)-g(r) \approx \frac{\sigma_TL}{4\pi r^2 \mu m_Hc}[M(r)-\frac{L_{\rm Edd}}{L}]=\frac{G}{ r^2}\mbh [M(r)\frac{L}{L_{\rm Edd}}-1]$. $L$ is the total luminosity, and approximately equals \lb. $M(r)$ is force multiplier, which is the ratio of the total radiation pressure to the radiation pressure due to Compton scattering. $M(r) \geq 1$, and  $M(r) = 1$ for fully ionized gas.  Assuming the \feii clouds falls down from torus (at $r_1$) with an initial $v_0=0$ to \feii region at $r_2$ ($\sim $ \rhb), for the cases of constant $[M(r)\frac{L}{L_{\rm Edd}}-1]$,  we get 
$v^2 =2 G \mbh (1-M \frac{L}{L_{\rm Edd}} )( \frac{1}{r_2}-\frac{1}{r_1})$. 
Assuming the size of the innermost part of the torus is $r_1=n r_2$, where $n\sim 3-4$ \citep{N2013}, and $\mbh=f\rm FWHM_{\hb}^2 r_2/G$, we get 
$\vfe^2 \approx 2\frac{n-1}{n} f {\rm FWHM^2_{\hb}}(1-M(r) \frac{L}{L_{\rm Edd}})$. The force multiplier $M(r)$ was found to be inversely correlated with the ionization parameter \citep{Chelouche2005}. Assuming $M(r) = M_0 (L/L_{\rm Edd})^{-\alpha}$, we get $\vfe^2 \approx 2\frac{n-1}{n} f {\rm FWHM^2_{\hb}}[1-M_0 (L/L_{\rm Edd})^{1-\alpha}]$.
A smaller $v$ would be found for AGN with larger \leddR, and smaller $\rm FWHM_{\hb}$ \citep[also see Figure 9 in ][]{Hu2008}. 



However, we find that it is not the case for the blueward shift of \feii, i.e., a smaller velocity of  \feii outflow for larger \mdot.  The reason for the relation between the blueward velocity \vfe and \mdot is not clear.

\subsection{The relation between $f$ and $\rm FWHM_{\hb}$}

\begin{figure*}
\center
\includegraphics[angle=0,width=6.5in]{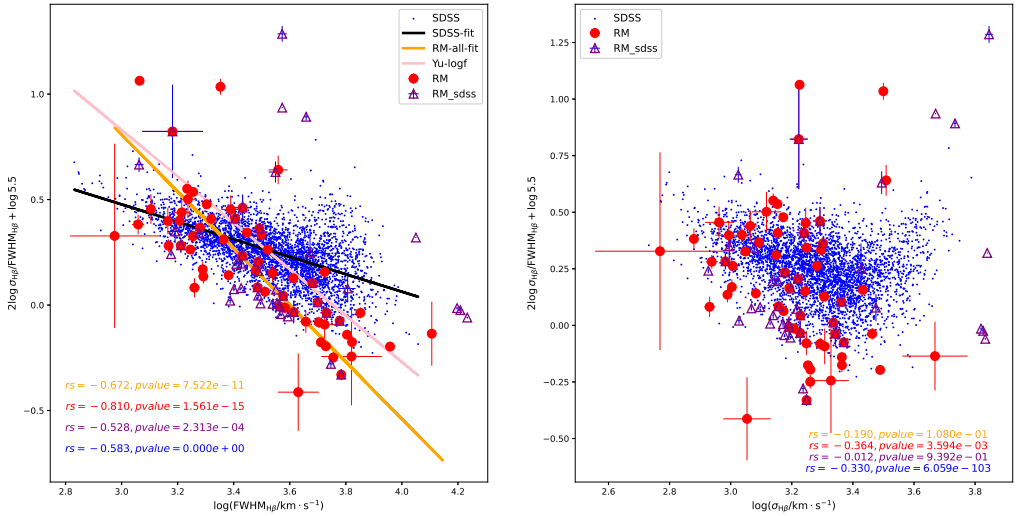}
\caption{Left: $f_{\rm FWHM}$ versus $\rm FWHM_{\hb}$. The black line and orange line are the best linear fittings for the SDSS subsample (blue points) and the RM subsample (red points and triangles). The triangles show 44 RM AGN from the SDSS. The red points show other 76 compiled RM AGN excluding 44 RM AGN from the SDSS. The pink line is the relation given by \cite{Yu2020b}. Right: $f_{\rm FWHM}$ versus \shb for the subsample of 120 RM AGN from the mean spectrum. The symbols are the same as in the left panel. The Spearman correlation coefficient and the probability of the null hypothesis are shown  in panels. The blue is for the SDSS Subsample, the  purple is for 44 RM AGN from the SDSS; the red is for the RM subsample excluding 44 RM AGN from the SDSS; the orange is for the RM subsample. }
\label{fig6}
\end{figure*}

Assuming a variable $f$ in $\rm FWHM_{\hb}$-based \mbh for the single-epoch or the mean spectrum, from Equation \ref{eq2}, $\log f=2\log (\rm \shb/FWHM_{\hb})+ log f_{\sigma} $. 
In the left panel in Figure \ref{fig6}, we show $\log f$ versus $\rm FWHM_{\hb}$ where $ f_{\sigma}=5.5$ as suggested by \cite{Yu2020b}.  For the SDSS subsample (blue points), the correlation is strong with $r_s=-0.583, p_{\rm null}=0$. Considering the errors in both coordinates, the best fitting is $\log f=(1.719\pm 0.028)-(0.414\pm 0.008)  \rm \log FWHM_{\hb}$ (black line in the left panel in Figure \ref{fig6}) for the single-epoch spectrum.  
For $\rm FWHM_{\hb}=1000~\kms, 2000 ~\kms, 6000 ~\kms, 14196 ~\kms$, $f=3, 2.25, 1.42, 1$, respectively. For the SDSS subsample, assuming the same \mbh from $\rm FWHM_{\hb}$, \shb and $f_{\sigma}=5.5$, the flatter slope means $f$ has a narrow range, i.e., from 3 to 1 for $\rm FWHM_{\hb}$ from 1000 to 14196 \kms .  For the value of $2.35$ for the single Gaussian profile, $\log f=2\log (\rm \shb/FWHM_{\hb})+ log f_{\sigma} =0.0017$.  The relation between $\log f$ and $\rm FWHM_{\hb}$ also shows the relation between $\rm FWHM_{\hb}/\shb$ and  $\rm FWHM_{\hb}$, $\log \rm FWHM_{\hb}/\shb=-0.49+0.207 \log \rm FWHM_{\hb}$.
In the left panel in Figure \ref{fig6}, almost all of the AGN with $\rm FWHM_{\hb} <4000~ \kms$  \citep[Population A,][]{Marziani2001}  in the SDSS subsample show  $\rm FWHM_{\hb}/\shb < 2.35$ deviated from that for the single Gaussian \hb profile.

For the RM subsample, the Spearman correlation test gives $r_s=-0.672, p_{\rm null}=7.522\times 10^{-11}$. Considering the errors in both coordinates, the best-fitting relation for RM subsample  shows $\log f =-(1.348\pm 0.080) \log \rm FWHM_{\hb}+(4.851\pm 0.27)$ ( orange solid line in Figure \ref{fig6}). 
Excluding 44  RM AGN from SDSS,  the correlation is stronger with $r_s=-0.81, p_{\rm null}=1.5\times 10^{-15}$. The best-fitting relation is  $\log f =-(1.357\pm 0.069) \log \rm FWHM_{\hb}+(4.883\pm 0.232)$. 
For 17 RM AGN with measured \sst \citep{Yu2020b}, it was found that the relation is  $\log f =-(1.10\pm 0.4) \log \rm FWHM_{\hb}+(4.13\pm 0.11)$ (pink solid line in the left panel in Figure \ref{fig6}). Although our fitting slope for the RM subsample is steeper than \cite{Yu2020b}, the trend is similar, i.e., larger $f$ for AGN with smaller $\rm FWHM_{\hb}$.  With our large number of AGN, the error of our slope is smaller, and our slope is consistent with the slope from RM AGN measured \sst \citep{Yu2020b}  considering their large error. 
For the SDSS subsample,  the slope derived from the single-epoch spectrum is flatter with respect to that for the RM subsample from the mean spectrum, i.e., $-0.414$ vs. $-1.348$. 

Using \sr as the benchmark instead of \sm in Equation \ref{eq2},  for 120 RM AGN, $r_s=-0.482, p_{\rm null}=9.18\times 10^{-4}$. Considering the errors in both coordinates, the best-fitting relation is  $\log f =-(1.418\pm 0.266) \log \rm FWHM_{\hb}+(5.148\pm 0.08)$.  Excluding 44  RM AGN from SDSS,  the correlation is stronger with $r_s=-0.616, p_{\rm null}=1.01\times 10^{-7}$. Considering the errors in both coordinates, the best-fitting relation is  $\log f =-(1.05\pm 0.40) \log \rm FWHM_{\hb}+(3.755\pm 0.40)$.  Considering their errors, these slopes are consistent with that using \sm as the benchmark. A smaller $\shb$-based $f_{\sigma}$ was suggested to be $4.47^{+1.41}_{-1.08}$ for the single-epoch spectrum \citep{Woo2015}, or $4.31\pm 1.05$ \citep{Grier2013} for the rms spectrum. 
Adopting these smaller $f$, it would lead our best fitting of the relation between $f$ and $\rm FWHM_{\hb}$  to have a smaller intercept by $\sim 0.1 ~\rm dex$. 

Because there is a strong relation between \mdot and $\rm FWHM_{\hb}$ \cite[e.g., ][]{Bian2003},  it implies that, using  $\rm FWHM_{\hb}$ as a velocity tracer in the \mbh calculation,  the corresponding $f$ has a correlation with \mdot, i.e., a larger $f$ is needed for AGN with larger \mdot. For a simple model of thick-disk BLRs, neglecting the contribution of outflow in the \hb profile, we obtained $f = 1/(a^2 + sin^2 \theta ) $, where  $a$ is the ratio of the scale height of the thick disk to the radius $r$, and $\theta$ is  the inclination of the thick-disk of BLRs to the line of sight \citep{Collin2006, Yu2019}. The cumulative fraction of this variable $f$ for 34 RM AGN with measured stellar velocity dispersion \sst was consistent with this simple model of thick-disk BLRs \citep{Yu2019}.  The relation between $f$ and $\rm FWHM_{\hb}$ implied that, as the tracer of BLRs velocity, $\rm FWHM_{\hb}$ has some dependence on the BLRs inclination, while \shb is insensitive to the inclination \citep{Collin2006, Yu2019}.

In right panel in Figure \ref{fig6}, we show $\log f$ versus \shb for the RM subsample (red points and triangles) and for the SDSS subsample (blue points). The Spearman correlation test gives $r_s=-0.190, p_{\rm null}=1.08\times 10^{-1}$ for the RM subsample, which is not too significant.  Excluding 44  RM AGN from SDSS, the correlation is always not strong with $r_s=-0.364, p_{\rm null}=3.59\times 10^{-3}$. Only for 44 RM AGN from SDSS,  the correlation is very weak with $r_s=-0.012, p_{\rm null}=9.392\times 10^{-1}$. For the SDSS subsample , $r_s=-0.33, p_{\rm null}=6.06\times 10^{-103}$, which is weaker than that with $\rm FWHM_{\hb}$. Using \sr as the benchmark instead of \sm  in Equation \ref{eq2} (see the right panel in Figure \ref{fig6}), the Spearman correlation test gives $r_s=-0.084, p_{\rm null}=5.88\times 10^{-1}$ for the RM subsample, which is not too significant.  Excluding 44  RM AGN from SDSS,  the correlation is always not strong with $r_s=-0.249, p_{\rm null}=5.09\times 10^{-2}$.  If assuming the virial factor $f$ has a dependence on the BLR inclination to the line of sight, thes not significant correlations in the right panel show that \shb has no dependence on the inclination. It is consistent with the constant $f$ using \shb as the velocity tracer in the \mbh calculation.

\section{Conclusions}
For a large sample of 4037 SDSS quasars ($z<0.8$) with measured \shb and \rfe and 120 compiled RM AGN,  we use \shb instead of $\rm FWHM_{\hb}$ and the extended \RL relation (including \rfe) to calculate \mbh from the single-epoch spectra for the SDSS subsample, and \shb from the mean spectra for the RM subsample.  We use \mdot instead of \leddR to indicate the accretion strength to investigate the relation with optical \feii emission (\rfe and \vfe). We also derive the relation between the virial  factor $f$ and $\rm FWHM_{\hb}$ for the single-epoch spectrum. 
The main conclusions can be summarized as follows:

\begin{itemize}
\item  For the  subsample of 4037 SDSS quasars, with respect to  $\rm FWHM_{\hb}$, $\shb$-based \mbh are  averagely larger by $0.25$ dex;  \mdot are  averagely smaller by $0.50$ dex.  There exists a nonlinear relationship between $\leddR$ and \mdot, $\leddR \propto \mdot^{0.52}$.  For the subsample of 120 RM AGN, the results are similar. 
With respect to  $\rm FWHM_{\hb}$, $\shb$-based \mbh (from the mean spectrum) for the RM subsample are  averagely larger by $0.23$ dex,  \mdot are  averagely smaller by $0.43$ dex, and $\leddR \propto \mdot^{0.56}$.  This relation of \leddR with \mdot can be derived assuming \lv is calculated from the standard accretion disk model and \lb is proportional to \lv. 

\item Adopting $\leddR \propto \mdot ^{0.5}\lv^{0.25}$, it means that the converting efficiency $\eta \propto \mdot^{-0.5}\lv ^{0.25} \propto (\leddR)^{-1}\lv^ {0.5}$. The nonlinear relation implies that $\eta$ is anticorrelated  with \mdot or \leddR (including a relatively weaker dependence on \lv), i.e., smaller efficiency with higher dimensionless accretion rate. The efficiency is related to the SMBH spin. 

\item For the  subsample of 4037 SDSS quasars,  a strong relation between \rfe and \mdot is found: $\rfe=-(0.936\pm0.006)+(0.600\pm 0.008)\log \mdot$ with the Spearman correlation coefficient $r_s$ of $0.727$. For the subsample of 120 RM AGN, the results are similar with a larger scatter; $\rfe=(0.713\pm0.047)+(0.532\pm 0.066)\log \mdot$ with $r_s=0.423$. With respect to \oiii ~$\lambda$5007, the velocity shift \vfe has an strong anticorrelation with \mdot, whenever inflow  ($r_s=0.612$) or outflow ($r_s=-0.419$). These strong correlations show that the optical \feii emission has an intimate connection with the SMBH accretion process. 

\item Assuming the \feii clouds fall down from torus with an initial $v_0=0$, we get the inflow velocity of the optical \feii
$\vfe^2 \propto  {\rm FWHM^2_{\hb}}(1-M(r) \frac{L}{L_{\rm Edd}})$.  A smaller $v$ would be found for AGN with larger \leddR, smaller $\rm FWHM_{\hb}$. It is qualitatively consistent with the relation of the inflow velocity \vfe with $\rm FWHM_{\hb}$ and \leddR or \mdot.

\item Assuming a variable $f$ in $\rm FWHM_{\hb}$-based \mbh from the single-epoch spectrum, we find a strong correlation between  $\log f$ and $\rm FWHM_{\hb}$, $\log f=(1.719\pm 0.028)-(0.414\pm 0.008)  \rm FWHM_{\hb}$.  It is flatter than that derived from the mean spectrum for the RM subsample or  17 RM AGN with measured \sst. However,  the relation between $\log f$ and  \shb is not too strong, suggesting that \shb doesn't seem dependent much on the BLR inclination and a constant $f_{\sigma}$ is suitable for \shb as the velocity tracer.

\end{itemize}

\acknowledgments

We are  very grateful to the anonymous referee for her/his instructive comments which significantly improved the content of the paper. 
This work has been supported by the National Science Foundations of China (Nos. 11973029 and 11873032). This work is supported by the National Key Research and Development Program of China (No. 2017YFA0402703).

\newpage

\begin{thebibliography}{}
\bibitem[Bentz et al. (2013)]{Be13} Bentz, M. C.,  et al. 2013, \apj, 767, 149
\bibitem[Bian \& Zhao (2003)]{Bian2003} Bian, W. H., \& Zhao, Y. H. 2003, PASJ, 55, 599
\bibitem[Bian \& Zhao (2004)]{Bian2004} Bian, W. H., \& Zhao, Y. H. 2004, MNRAS, 347, 607
\bibitem[Bian et al. (2008)]{Bian2008} Bian, W. H., et al. 2008, MNRAS, 390, 752
\bibitem[Blandford \& McKee (1982)]{BM82} Blandford, R., McKee, C. 1982, \apj, 255, 419
\bibitem[Boroson \& Green (1992)]{BG92}  Boroson, T. A., \& Green, R. F., 1992, \apjs, 80, 109
\bibitem[Boroson(2002)]{Boroson2002} Boroson, T. A. 2002, ApJ, 565, 78
\bibitem[Chelouche \& Netzr (2005)]{Chelouche2005} Chelouche, D., \& Netzer, H. 2005, \apj, 625, 95
\bibitem[Collin et al. (2002)]{Collin2002} Collin, S., et al. 2006, A\&A, 388, 771
\bibitem[Collin et al. (2006)]{Collin2006} Collin, S., Kawaguchi, T., Peterson, B. M., \& Vestergaard, M. 2006, A\&A,
456, 75
\bibitem[Dalla Bont\`{a} et al. (2020)] {Dalla2020} Dalla Bont\`{a}, E., Peterson B. M., Bentz, M. C., et al. 2020, \apj, 903, 112

\bibitem[Davis \& Laor (2011)]{DL2011}Davis, S. W., \& Laor, A. 2011, \apj, 728, 98
\bibitem[Du et al. (2016a)]{Du2016a} Du, P., et al. 2016a, \apj, 818, L14
\bibitem[Du et al. (2016b)]{Du2016b} Du, P., et al. 2016b, \apj, 825, 126
\bibitem[Du \& Wang (2019)]{Du2019} Du, P., \& Wang, J.-M. 2019, \apj, 886, 42
\bibitem[Ge et al. (2016)]{Ge2016} Ge, X., Bian, W. H., Jiang, X. L., et al., 2016, MNRAS, 462, 966
\bibitem[Grier et al. (2013)]{Grier2013} Grier, C. J., Martini, P., Watson, L. C., et al. 2013, ApJ, 773, 90
\bibitem[Grier et al. (2017)]{Grier2017} Grier, C. J., Trump, J. R., Shen Y., et al. 2017, \apj, 851, 21
\bibitem[Ho \& Kim (2014)]{HK14} Ho, L., \& Kim, M. 2014, \apj, 789, 17
\bibitem[Hu et al. (2008)]{Hu2008} Hu, C., Wang, J.-M., Ho, L. C., et al. 2008, ApJ, 687, 78
\bibitem[Hu et al. (2012)]{Hu2012} Hu, C., Wang, J.-M., Ho, L. C., et al. 2012, ApJ, 760, 126
\bibitem[Hu et al. (2015)]{Hu2015} Hu, C., Du, P., Lu, K. X., et al. 2015, ApJ, 804, 138
\bibitem[Hu et al. (2021)]{Hu2021} Hu, C., Li, S. S., Yang, S.,  et al. 2021, ApJS, 253, 20
\bibitem[Huang et al. (2020)]{Huang2020} Huang, J. et al. 2022, ApJ, 895, 114
\bibitem[Kaspi et al. (2000)]{Ka00} Kaspi, S., et al. 2000, \apj, 533, 631
\bibitem[Kilerci Eser et al. (2015)]{KE15} Kilerci Eser, E., Vestergaard, M., Peterson, B. M., Denney, K. D., Bentz, M. C. 2015, \apj, 801, 8
\bibitem[Khadka et al. (2022)]{Khadka2022} Khadka, N., Mart{\'\i}nez-Aldama, M. L., Zaja\v{c}ek, M., et al. 2022, MNRAS, 513, 1985

\bibitem[Maithil et al. (2022)]{Maithil2022} Maithil J., Brotherton, M. S., Shemmer, O., et al. \mnras, in press, (arXiv:2206.11486) 

\bibitem[Marconi et al. (2004)]{Marconi2004} Marconi, A., Risaliti, G., Gilli, R. et al. 2004, MNRAS, 351, 169
\bibitem[Mart{\'\i}nez-Aldama et al.(2019)]{Ma2019} Mart{\'\i}nez-Aldama, M. L., Czerny, B., Kawka, D., et al. 2019, 883, 170 

\bibitem[Marziani et al. (2001)]{Marziani2001} Marziani, P., Sulentic, J. W., Zwitter, T., et al. 2001, ApJ, 558, 553
\bibitem[Mej{\'\i}a-Restrepo et al. (2018)]{Mejia2018} Mej{\'\i}a-Restrepo, J. E., et al. 2018, Nature Astronomy, 2, 63  
\bibitem[McGill et al. (2008)]{McGill2008} McGill, K. L., Woo, J.-H., Treu, T., \& Malkan, M. A. 2008, ApJ, 673, 703
\bibitem[Netzer \& Trakhtenbrot (2007)]{N2007} Netzer, H., \& Trakhtenbrot, B. 2007, ApJ, 654, 754
\bibitem[Netzer (2013)]{N2013} Netzer, H. 2013, The Physics and Evolution of Active Galactic Nuclei
\bibitem[Onken et al. (2004)]{On04} Onken, C. A., et al. 2004, \apj, 615, 645
\bibitem[Panda (2022)]{Panda2022} Panda,  S. 2022, Front. Astron. Space Sci., 9, 850409
\bibitem[Peterson et al. (2004)]{Pe04} Peterson, B. M., et al. 2004,  \apj, 613, 682
\bibitem[Shakura \& Sunyaev (1973)]{SS73}Shakura, N. I., \& Sunyaev, R. A. 1973, A\&A, 24, 337
\bibitem[Shen et al. (2011)]{Sh11} Shen, Y., et al. 2011, \apjs, 194, 45
\bibitem[Shen \& Ho (2014)]{Shen2014} Shen, Y., \&  Ho, L. C. 2014, Nature, 513, 210
\bibitem[Shen et al. (2019)]{Shen2019} Shen, Y., Hall, P. B., Horne, K., et al. 2019, ApJS, 241, 34
\bibitem[Sulentic et al. (2012)]{Sulentic2012} Sulentic, J. W., Marziani, P., Zamfir, S., \& Meadows, Z. A. 2012, ApJ, 752, L7
\bibitem[Urry \& Padovani (1995)]{Urry1995} Urry, C. M., \& Padovani, P. 1995, \pasp, 107, 803
\bibitem[Wang et al. (2013)]{Wang2013} Wang, J. M., Du, P., Valls-Gabaud, D., Hu, C., Netzer, H., 2013, Phys. Rev.
Lett., 110, 081301
\bibitem[Wang et al. (2014)]{Wang2014}Wang, J. M., Du P., Li Y.-R., Ho L. C., Hu C., Bai J.-M., 2014, \apj, 792, L13
\bibitem[Woo et al. (2015)]{Woo2015} Woo, J. H., et al. 2015, \apj, 801, 38
\bibitem[Yu et al. (2019)]{Yu2019} Yu, L. M., Wang, C., Bian, W. H., Zhao, B. X.,  Ge, X. 2019, \mnras, 488. 1519
\bibitem[Yu et al. (2020a)]{Yu2020a} Yu, L. M., Zhao, B. X., Bian W. H., et al. 2020, \mnras, 491, 5881
\bibitem[Yu et al. (2020b)]{Yu2020b} Yu, L. M., Bian, W. H., Zhang, X. G., et al. 2020, \apj, 901, 133
\bibitem[Zhao et al. (2020)]{Zhao2020} Zhao, B. X., Bian, W. H., Yu, L. M., Wang, C. 2020, ApSS, 365, 22
\end{thebibliography}

\end{document}